\documentclass[fleqn,usenatbib]{mnras}

\usepackage{newtxtext,newtxmath}

\usepackage[T1]{fontenc}

\DeclareRobustCommand{\VAN}[3]{#2}
\let\VANthebibliography\thebibliography
\def\thebibliography{\DeclareRobustCommand{\VAN}[3]{##3}\VANthebibliography}

\usepackage{graphicx}	
\usepackage{amsmath}	
\usepackage{multirow}
\usepackage{url}
\usepackage{xurl}
\usepackage{makecell}

\newcommand{\kepler}{\textit{Kepler}}

\newcommand{\gaia}{\textit{Gaia}}
\newcommand{\jwst}{\textit{JWST}}
\newcommand{\hst}{\textit{HST}}

\newcommand{\galex}{\textit{GALEX}}

\newcommand{\tess}{\textit{TESS}}
\newcommand{\swift}{\textit{Swift}}

\newcommand{\Msun}{$\mathrm{M_{\odot}}$}
\newcommand{\fuvonethirty}{$\mathrm{FUV_{130}}$}

\newcommand{\kms}{$\mathrm{kms^{-1}}$}
\newcommand{\wlonethirty}{$\mathrm{WL_{130}}$}

\title[Testing flare models with \tess\ and \hst]
{Optically Quiet, But FUV Loud: Results from comparing the far-ultraviolet predictions of flare models with \tess\ and \hst}

\author[J. A. G. Jackman et al.]{
James A. G. Jackman $^{1}$\thanks{E-mail: jamesjackmanastro@gmail.com},
Evgenya L. Shkolnik $^{1}$,
R. O. Parke Loyd $^{2}$,
Tyler Richey-Yowell \thanks{Percival Lowell Postdoctoral Fellow} $^{3}$,
\\
$^{1}$School of Earth and Space Exploration, Arizona State University, Tempe, AZ, 85287, USA\\
$^{2}$Eureka Scientific, 2452 Delmer Street Suite 100, Oakland, CA 94602-3017, USA\\
$^{3}$Lowell Observatory, 1400 W Mars Hill Rd, Flagstaff, AZ, 86001, USA\\
}

\date{Accepted XXX. Received YYY; in original form ZZZ}

\pubyear{2022}

\begin{document}
\label{firstpage}
\pagerange{\pageref{firstpage}--\pageref{lastpage}}
\maketitle

\begin{abstract}
The far-ultraviolet (FUV) flare activity of low-mass stars has become a focus in our understanding of the exoplanet atmospheres and how they evolve. However, direct detection of FUV flares and measurements of their energies and rates are limited by the need for space-based observations. The difficulty of obtaining such observations may push some works to use widely available optical data to calibrate multi-wavelength spectral models that describe UV and optical flare emission. These models either use single temperature blackbody curves to describe this emission, or combine a blackbody curve with archival spectra. These calibrated models would then be used to predict the FUV flare rates of low-mass stars of interest. To aid these works, we used \tess\ optical photometry and archival \hst\ FUV spectroscopy to test the FUV predictions of literature flare models. We tested models for partially (M0-M2) and fully convective (M4-M5) stars, 40 Myr and field age stars, and optically quiet stars. We calculated FUV energy correction factors that can be used to bring the FUV predictions of tested models in line with observations. \textcolor{black}{A flare model combining optical and NUV blackbody emission with FUV emission based on \hst\ observations provided the best estimate of FUV flare activity, where others underestimated the emission at all ages, masses and activity levels, by up to a factor of 104 for combined FUV continuum and line emission and greater for individual emission lines. }
We also confirmed previous findings that showed optically quiet low-mass stars exhibit regular FUV flares. 

\end{abstract}




\section{Introduction} \label{sec:intro}
The ultraviolet (UV) emission of stellar flares has been highlighted in recent years for its role in exoplanet habitability and the evolution of exoplanetary atmospheres. Far-ultraviolet (FUV, 1150-1700\AA) photons can photodissociate $\mathrm{H_{2}O}$, $\mathrm{CO_{2}}$ and alter the concentration of atmospheric HCN, \textcolor{black}{a species considered to be an important} biosignature \citep[][]{Rimmer19}. Planets subjected to regular FUV flare emission may also have permanently altered atmospheres, as they are unable to return to their initial state between regular flare events \citep[e.g. requiring $>$30 years for species to return to their quiescent states][]{Venot16}.  
The ratio of the FUV to the near-UV (NUV) flux from a star also contributes to the rate at which atmospheric hazes form \citep[e.g.][]{Arney17}. 
These hazes can complicate searches for biosignatures in exoplanet atmospheres via transmission spectroscopy with instruments such as \jwst. Therefore, in order to understand the atmosphere of exoplanets around low-mass stars, in terms of both the evolution and the context of observed signals, we must understand the FUV activity of their host stars. 

Stellar flares occur when a magnetic field lines in the outer atmosphere of a star undergo a magnetic reconnection event \citep[e.g.][]{Benz10}. The reconnection event releases energy into the surrounding coronal plasma and accelerates charged particles downwards along magnetic field lines towards the chromosphere and photosphere, where field loops are anchored. 
These nonthermal particles impact the dense chromospheric plasma and are rapidly decelerated, releasing their energy as they do so \citep[e.g.][]{Hudson72}. This energy release heats and evaporates the surrounding plasma at the footpoints. The evaporated plasma rises to fill the newly reconnected magnetic field loop and emits in X-rays. At the same time, a dense heated layer descends towards the lower chromosphere and photosphere \citep[][]{Kowalski18}. This layer, termed a chromospheric condensation, cools as it descends and in turn heats lower layers. High energy nonthermal particles are also thought in some cases to directly heat the lower chromosphere and photosphere, however this may depend on the local plasma conditions and energy deposition rate \citep[e.g.][]{Watanabe17,Watanabe20}. 
The footpoint emission gives rise to the optical signatures of stellar flares, which are regularly detected with wide-field exoplanet surveys such as \kepler, \tess\ and NGTS \citep[][]{Borucki10,Ricker14,Wheatley18}. Studies often use these optical observations as a tracer for the flare activity in other wavelengths, including the UV \citep[e.g.][]{Feinstein20,Chen21}. 

In order to extrapolate optical observations into the UV, studies must make assumptions about the flare spectrum and the correlation between UV and optical flare emission. Models range from a blackbody spectrum with a fixed temperature \citep[][]{Shibayama13} to the use of archival spectra covering FUV to optical wavelengths that include time-dependent line emission and continuum features \citep[][]{Segura10,Tilley19}. These models have been used to constrain the contribution of UV flare emission to ozone depletion, atmospheric evaporation and even the formation of chemicals required for life to emerge \citep[][]{Tilley19,Feinstein20,Rimmer18}.  These studies typically assume that every optical flare has a UV counterpart, and vice versa, providing a way to predict the UV flare activity of a star from optical data alone. 

Recent works have called into question the accuracy of current flare models and the assumptions used to predict the UV flare emission from optical data. 
\citet{Kowalski19} found that the Balmer continuum and line emission increased the NUV flux to a factor of 2--3 above the prediction of an extrapolated optical continuum. \citet{Jackman23} and \citet{Brasseur23} compared archival \galex\ photometry to optical observations (\tess\ for M stars and \kepler\ for FGK stars respectively), finding that flares emitted greater amounts of NUV emission than expected from extrapolation of optical emission. 
\citet{Brasseur23} identified the presence of short duration NUV flares detected with \galex\ that lacked detectable optical counterparts in simultaneous \kepler\ long and short cadence data. These results were taken as evidence that the NUV emission from flares was greater than models calibrated using optical observations would predict. 

Studies in the FUV have also found discrepancies between prediction and observation, \textcolor{black}{to a greater extent than in the NUV}. \textcolor{black}{As part of the HAZMAT programme \citep[][]{Shkolnik14}, \citet{Miles17} used archival \galex\ observations to measure the NUV and FUV variability of low-mass stars. They found evidence that flares increased the FUV flux density by a greater amount up to 4.6 times than in the \galex\ NUV bandpass.}  \citet{Jackman23} found that existing flare models calibrated using \tess\ observations were unable to predict the \galex\ NUV and FUV flare rates of field age M stars. Models underestimated the \galex\ NUV emission of fully convective M stars by at least a factor of 6.5, and the \galex\ FUV flare emission by \textcolor{black}{30.6}. 
FUV observations of flares from Proxima Centauri with \hst\ have shown a spike in the FUV flux that precedes the bulk of the white-light emission, suggesting the FUV also arises from the initial heating of the chromosphere, possibly from the intensely heated chromospheric condensation \citep[][]{MacGregor21}. \textcolor{black}{The disconnect between the optical and FUV flare emission has been highlighted in multiple studies. \citet{France13} and \citet{France16} noted that optically inactive M stars in the MUSCLES program showed FUV flaring activity. Subsequent studies took this further and focused on individual stars. }
\citet{Loyd20} used archival \textit{Hubble Space Telescope} (\hst) observations \textcolor{black}{from the MUSCLES programme \citep[][]{France13, France16}} to identify FUV flares from GJ 887, which showed no sign of magnetic activity (flares, starspots) in its \tess\ optical lightcurve. \citet{Loyd23} identified similar results for the optically quiet M star GJ 436, detecting 14 FUV flares in 21.12 hours of cumulative exposure. \citet{France20} also found frequent flares from the 10 Gyr Barnard's star (GJ 699), detecting two flares in 3.6 hours of observation and estimating a flare duty cycle of $\approx$ 25 per cent. 
These observations highlighted that some FUV flares may occur without detectable optical signatures, and stars that were previously considered magnetically inactive may exhibit regular FUV flaring behaviour. FUV spectroscopy \textcolor{black}{taken during the HAZMAT \citep[][]{Shkolnik14} and MUSCLES programmes} has also been used to isolate pseudo-continuum emission and measure flare temperatures of 15,500\,K and up to 40,000\,K \citep[][]{Loyd18hazmat,Froning19}. These values are far above the 9000\,K value typically assumed by optical flare studies. Multi-colour optical photometry of flares have also measured flare temperatures up to 40,000\,K \citep[][]{Howard20}. 

The FUV discrepancy is a clear issue in multi-wavelength flare models and the studies that use them. However, 
testing models for individual flares with simultaneous observations is a difficult task. The timing of an individual flare can not be predicted, and the contrast between the flare and quiescent spectrum is greatly reduced in the optical relative to the FUV. In order to detect multiple events in both wavelengths, such that energy ratios and occurrences can be calculated and compared, continuous observations over tens of hours are required. In addition to this, FUV observations can currently only be obtained with \hst. As flares can increase the FUV brightness of low-mass stars by factors of thousands, brightness limits are in place for flare stars in order to protect detectors for \hst\ instruments \citep[][]{Osten17cos,Osten17stis}.

An alternative approach is to use archival data to study averaged FUV and optical flare emission. By studying stars that had been observed in the UV and optical at different times, inferences can be made about the nature of their flare emission. As mentioned above, \citet{Brasseur23} and \citet{Jackman23} used this approach to study the \galex\ UV and optical emission of stars, using \kepler\ and \tess\ respectively. The availability of optical lightcurves from long baseline, wide-field exoplanet surveys means we can utilise archival UV data from most of the sky. While this approach cannot tell us about how the relative fluxes change throughout individual flares, it can shed light on energy partitions, relative flare rates and how these might change with spectral type and age. These results are used to test the UV predictions of flare models, in turn improving the accuracy of future exoplanet habitability studies. 

In this work we tested the FUV predictions of flare models for low-mass stars. 
We combined \tess\ optical observations with archival FUV TIME-TAG spectroscopy from \hst\ to study how the FUV emission of flares is underestimated by models extrapolated from the optical. \citet{Jackman23} used \galex\ photometry in their work, which limited their work to tests of the total UV flux within the \galex\ NUV and FUV bands. By using FUV spectroscopy we are able to isolate continuum and line emission, providing more detail about how well models can reproduce the flux associated with these features.  

In Sect.\,\ref{sec:data} we detail the instruments, targets and data used in this work.  
In Sect.\,\ref{sec:methods} we describe how we detected flares in each dataset and tested the FUV predictions of flare models. In Sect.\,\ref{sec:results} we present the results of our model testing and in Sect.\,\ref{sec:discussion} we discuss the impact of our results and how future studies can test flare models further.

\section{Data} \label{sec:data}
In this section we discuss the data we used for the optical and ultraviolet analysis in this work. 

\subsection{\hst}
We used the \hst\ archive to identify low-mass stars that had previously been observed with either the \textit{Cosmic Origins Spectrograph} (COS) or the \textit{Space Telescope Imaging Spectrograph} (STIS). These two instruments are used onboard \hst\ for FUV spectroscopy and have both been used to study the FUV emission of stellar flares \citep[e.g.][]{Hawley03,Loyd18muscles,MacGregor21}. 

We filtered our search for observations that had been taken in the TIME-TAG mode. In the TIME-TAG mode the arrival time and the position on the detector are recorded for every arriving photon. This is done with a typical precision of 125 microseconds, allowing the construction of lightcurves with user-defined cadences. We downloaded data taken with the COS G130M grating, and the STIS E140M and G140L gratings. We chose these gratings as they cover a range of chromospheric and transition region emission lines frequently used to study the FUV emission and activity of low-mass stars \citep[e.g.][]{Loyd18muscles,DiamondLowe21,Feinstein22}. Table.\,\ref{tab:line_emission} shows the emission lines and continuum regions used in this work.

\subsubsection{\hst\ Lightcurve Generation} \label{sec:data_hst_lc_generation}
We generated FUV lightcurves from archival \hst\ COS and STIS datasets.  
FUV lightcurves were generated from COS G130M datasets following the method outlined by \citet{DiamondLowe21}. This method is itself based on one described by \citet{Loyd14} and expanded upon by \citet{Loyd18muscles}. We used the COS \textsc{corrtag} TIME-TAG data files to generate lightcurves in our analysis. The \textsc{corrtag} files contain the photon arrival position on the detector, along with wavelength and the recorded time of arrival for each photon. This information enables us to portion data into time bins, and create lightcurves from the data with a user-specified cadence. For data in a given time bin, we first measured the gross spectrum for the source region (the source plus the background), and the background alone in counts. We calculated the uncertainty for both of these, assuming a Poisson distribution. \textcolor{black}{We noted in our analysis that the pipeline-provided uncertainties on the flux-calibrated spectra for faint sources could result in negative fluxes within 3$\sigma$, impacting our integrated fluxes when generating lightcurves. To account for this we propagated Poisson uncertainties from the raw counts in the source and background spectra into the integrated lightcurves.} We measured the background spectrum for each source following the method detailed in the COS data handbook \footnote{\url{https://hst-docs.stsci.edu/cosdhb/chapter-3-cos-calibration/3-4-descriptions-of-spectroscopic-calibration-steps}}. We used the same background regions as those defined in the COS \textcolor{black}{TWOZONE} references files and used in the CALCOS data reduction pipeline. We used the associated x1d spectrum for the full exposure, calculated by the CALCOS pipeline, to flux calibrate our source and background spectra \citep[e.g.][]{Loyd18muscles}. 

In order to propagate our Poisson uncertainties \textcolor{black}{from the raw counts to the integrated lightcurves}, we used a Monte Carlo process. For a given spectrum, we randomly generated a series of synthetic spectra with values drawn from Poisson distributions. We did this for both the measured source and background spectra. We subtracted the randomly generated background spectra from the randomly generated gross source spectra, to produce a sample of source-only spectra. We applied the measured flux calibration to these resulting net counts spectra. Each flux calibrated spectrum was integrated over to produce lightcurves. We took the median as our measured value and the 16th and 84th percentiles were used to calculate the lower and upper uncertainties. \textcolor{black}{This Monte Carlo process accounted for the different lower and upper uncertainties in the detected counts and resulted in asymmetrical uncertainties in our integrated lightcurves, something important for the fainter sources in our sample.}  

We generated FUV lightcurves from STIS TIME-TAG data by following a similar method to the one described by \citet{Loyd14}. We extracted the signal and performed a background subtraction for each order in the STIS echelle spectra. We obtained the positions of the spectrum trace and background regions, along with the size of the extraction regions for both from the reduced X1D spectrum data for each \hst\ visit. We performed the background subtraction following the method outlined in the STIS data handbook. We performed the data reduction ourselves as we noted early on in our analysis that the 1D spectra for some sources reduced by CALSTIS had negative gross flux values in the continuum regions. By reducing the data ourselves we were able to avoid this issue, and like for the COS data were able to propagate Poisson uncertainties for both the source and background measurements into our final measured lightcurves. 

We used a fixed cadence of 30 seconds when generating FUV lightcurves. This value was chosen to resolve flares and potential substructure without compromising on the signal to noise ratio \citep[e.g.][]{Million16,Feinstein22}. We chose a fixed cadence for all lightcurves to provide consistency across our dataset and to simplify flare injection and recovery tests performed in Sect.\,\ref{sec:method_testing_models}. 

\begin{figure}
    \centering
    \includegraphics[width=\columnwidth]{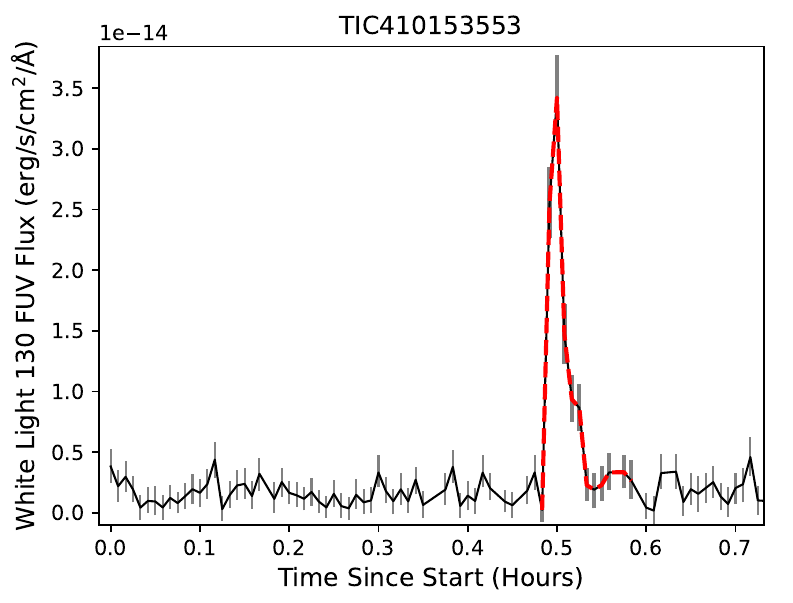}
    \caption{\textcolor{black}{Example of a lightcurve generated from \hst\ TIME-TAG spectra with a cadence of 30 seconds. This source had an average signal to noise ratio of 1.5 in the white-light 130 bandpass, but this increased to 9.7 at the peak of the flare. The red dashed region of the lightcurve indicates a detected flare.}}
    \label{fig:flare_1_5}
\end{figure}

\begin{figure}
    \centering
    \includegraphics[width=\columnwidth]{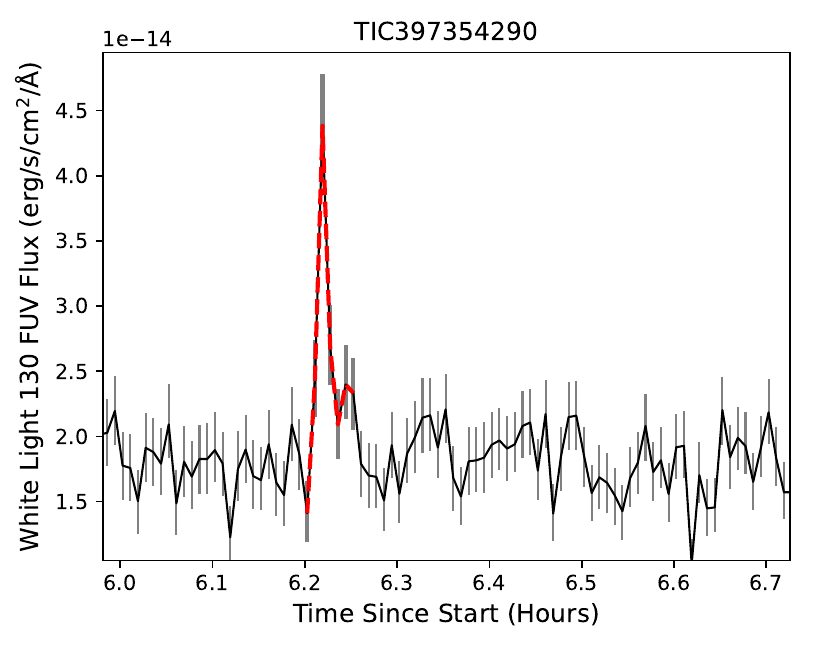}
    \caption{\textcolor{black}{Example of a lightcurve with an average signal to noise ratio of 8.8 in the white-light 130 bandpass, with a value of 11.1 at the flare peak. Note that the higher signal to noise ratio of the quiescent light curve enables the detection of lower amplitude flares, with this event having a flare amplitude of 1.7, compared to a value of 12 for the flare shown in \ref{fig:flare_1_5}.}}
    \label{fig:flare_8_8}
\end{figure}

We created \textcolor{black}{three} sets of lightcurves for each star in our sample. The integration regions 
and wavelengths corresponding to different lightcurves are given in Tab.\,\ref{tab:line_emission}. The first set of lightcurves were white-light lightcurves for the entire dataset. These integrated over both line and continuum emission in the regions covered by the COS and STIS spectrographs. We designed our integration regions to be similar to those used in the \fuvonethirty\ lightcurves defined by \citet{Loyd18muscles}. These regions include both the FUV continuum and line emission but mask Ly$\alpha$, O I and emission due to the effects of  geocoronal airglow. 
We note that the white-light lightcurves associated with the G130M grating do not go beyond 1362.7\AA\ (unlike the \fuvonethirty\ lightcurves) due to the presence of COS G130M datasets using central wavelengths shorter than 1291\AA\ in our sample. We also generated \fuvonethirty\ lightcurves for stars that were observed with the STIS E140M and the COS G130M gratings, specifically the COS G130M 1291, 1300, 1318 and 1327\AA\ central wavelengths. These lightcurves cover a subset of our sample, but by generating them we can compare our testing of flare models with previous works.  
 
The second set of lightcurves were pseudo-continuum lightcurves. These used integration regions that aimed to avoid strong line emission, allowing us to test the continuum emission predictions of flare models. We used the CHIANTI database to identify wavelength regions without strong line emission \citep[][]{chianti}. We note that these lightcurves may still have some contribution from low-level emission lines, \textcolor{black}{and may be dominated by}, low-level emission lines that are only resolved in the highest S/N FUV datasets \citep[e.g.][]{MacGregor21, Feinstein22} hence us using the term pseudo-continuum to describe them \citep[e.g.][]{Jackman23}. 
 
The third set of lightcurves isolated the line emission. The lines we used are given in Tab.\,\ref{tab:line_emission}. These lines trace the transition region and the chromosphere, and have been the subject of various flare studies. The central wavelengths of each emission line are given in Tab.\,\ref{tab:line_emission}. We calculated the integration regions for each line by calculating the wavelengths 100\,\kms\ either side \citep[e.g.][]{Loyd18muscles}. This was done to include emission from the line wings. Overlapping integration regions were merged during our analysis.

\renewcommand{\cellalign}{l}
\begin{table}
    \centering
    \begin{tabular}{l|l}
         Name & Wavelength (\AA) \\
         \hline
         \makecell{White-light 130 \\(G130M, E140M, G140L)} & \makecell{1173.65--1198.49, 1201.71--1206.50,\\ 1223.00--1273.50, 1328.20--1354.49, \\ 1356.71--1357.59, 1359.51--1362.70} \\ \hline
         \fuvonethirty & \makecell{1173.65--1198.49, 1201.71--1212.16, \\ 1219.18--1274.04, 1329.25--1354.49, \\ 1356.71--1357.59, 1359.51--1428.90}
         \\ 
         \hline
         Pseudo-continuum 130 & \makecell{1173.65--1174.50, 1176.80--1190.00, \\ 1223.00--1238.40, 1239.30--1242.00, \\ 1243.50--1273.50, 1329.00--1334.00, \\ 1336.00--1354.49, 1356.71--1357.59, \\ 1359.51--1362.70}\\
         \hline
         Si IV & 1393.76, 1402.77 \\
         Si III & 1206.51 \\
         C II & 1334.53, 1335.71 \\
         C III & \makecell{1174.93, 1175.25, 1175.59, \\ 1175.71, 1175.99, 1176.37} \\
         N V & 1238.82, 1242.80 \\
         \hline
         
    \end{tabular}
    \caption{The lines and continuum integration regions}
    \label{tab:line_emission}
\end{table}

\begin{table*}
    \centering
    \begin{tabular}{l|c|c|c|c|c|c|c|c|c|c}
     Name & $\mathrm{WL_{130}}$ &
     \fuvonethirty\ & $\mathrm{PC_{130}}$ & Si IV & Si III & C II & C III & N V \\
     \hline
     Partially Convective Field Age M Stars & 17 & 15 & 17 & 15 & 14 & 17 & 17 & 16 \\
     Fully Convective Field Age M Stars & 6 & 6 & 6 & 6 & 6 & 6 & 6 & 6 \\
     Field Age (Tuc-Hor comparison) & 24 & 22 & 24 & 22 & 21 & 24 & 24 & 23  \\
     Tuc-Hor M & 12 & 12 & 12 & 12 & 12 & 12 & 12 & 12  \\
     Optically Quiet Field Age & 17 & 15 & 15 & 16 & 17 & 17 & 17 & 16 \\

    \end{tabular}
    \caption{The number of stars in each subset. Differences are due to the variety of gratings used to observe each star. \textcolor{black}{There is some overlap between samples, with 13 stars being present in both the partially convective and optically quiet field age samples, and 4 between the fully convective and optically quiet field age samples. }}
    \label{tab:stellar_selection}
\end{table*}

\subsection{\tess}
The \textit{Transiting Exoplanet Survey Satellite} (\tess) is a wide-field photometric survey telescope \citep[][]{Ricker14}. It observes individual 24$\times$96 square degree sectors of sky. Each sector is split into four, with each region being observed by one of the four \tess\ cameras. A single sector is observed for a duration of 27.4 days. \tess\ began its primary mission in July 2018 and observed the southern and northern hemisphere for one year each, with the primary mission ending in July 2020. \tess\ has been commissioned for two extended missions. The first of these lasted until September 2022 and the second is ongoing at the time of writing. The primary mission observed stars at two cadences. Full frame images were obtained with a cadence of 30 minutes and postage stamps of pre-selected stars were observed with a cadence of 2 minutes. An additional 20 second cadence mode was added in the first extended mission, while the cadence of the full frame images was changed to 10 minutes. 
2-minute and 20-second cadence lightcurves are automatically generated by the Science Processing Operations Center pipeline \citep[SPOC;][]{Jenkins16}, and are available for download from MAST. We elected to use the 2-minute cadence lightcurves in this work. We did this for consistency in the observation cadence between \tess\ sectors from the primary and extended missions, something that proved important during our flare injection and recovery in Sect.\,\ref{sec:method_tess}. The \tess\ lightcurves are provided in two types. These are the Simple Aperture Photometry (SAP) and Pre-Search Data Conditioned SAP (PDC\_SAP) lightcurves. Long term trends and variations have been removed from the PDC\_SAP lightcurves, but shorter timescale variations such as those from flares and transits have been left. We used the Pre-Search Data Conditioned (PDC\_SAP) SPOC lightcurves in our analysis. 

\subsection{Stellar Sample} \label{sec:method_sample}
In order to analyse how the FUV accuracy of optically-calibrated flare models changes with age and mass, we compiled samples of stars of young and old ages that had both \hst\ FUV and \tess\ optical observations. 
The number of stars in each sample and those that were used in each FUV analysis are given in Tab.\,\ref{tab:stellar_selection}.
 
The first two sets of these were field age partially and fully convective M stars. 
We selected these stars following the method outlined by \citet{Jackman23}. We calculated the absolute \gaia\ $M_{G}$ magnitude of the stars in our sample and compared these to the median fiducial curve from \citet{Jackman21}. We identified stars which fell between $\Delta M_{G}=+0.65$ and $-0.55$ of the median stellar magnitude as residing on the main sequence. We then used the listed masses from the \tess\ Input Catalog to filter this sample into partially and fully convective field age M stars. We followed \citet{Jackman23} and used mass ranges of 0.37--0.6\,\Msun\ (M0-M2) and 0.1--0.29\,\Msun\ (M4-M5) for the partially and fully convective samples respectively \citep[][]{Cifuentes20}. The interior structure of low-mass stars changes from partially to fully convective at around 0.3--0.35\Msun \citep[e.g.][]{Chabrier97,Macdonald18}. These mass ranges were chosen as they sample both sides of the transition in interior structure. 
We identified \textcolor{black}{17} and \textcolor{black}{6} stars in the partially and fully convective subsets. 

The next set we used in our analysis were  
40 Myr old stars associated with the Tuc-Hor moving group. These stars were observed with \hst\ COS with the G130M grating, as part of the HAZMAT programme (program ID 14784, PI Shkolnik). These observations were analysed by \citet{Loyd18hazmat}, who identified that when expressed in relative units, the FUV flare rate of low-mass stars does not change between 40 Myr and field ages. When expressed in terms of emitted energy, the younger stars showed flares with energies 10-100x that of the field age sample. We used these stars in our analysis to test whether the FUV accuracy of flare models of young stars relative to their field age counterparts, something that might point towards a change in FUV flare spectra with age. We identified 12 stars that had both \hst\ and \tess\ observations. 

The Tuc-Hor group used by \citet{Loyd18hazmat} had spectral types between M0.0 and M2.3 \citep[][]{Kraus14}. At 40 Myr this spectral range corresponds to a mass range of 0.35 and 0.65 \Msun \citep[][]{Pecaut13, Baraffe15}. Stars in this mass range undergo an increase in their surface temperature as they contract and move along the Hayashi track towards the main sequence, changing the range of observed spectral types. To compare flare model accuracy at 40 Myr and field ages we selected a separate sample of field age stars in the same mass range as the Tuc-Hor sample. We identified a sample of \textcolor{black}{24} field age stars that sit in the same mass range as the Tuc-Hor sample. We used these to probe how model accuracy changes with age in Sect.\,\ref{sec:discussion_age}. 

We used a subset of our field age low-mass stars to study the FUV activity of optically quiet low-mass stars. There were 17 stars with masses between \textcolor{black}{0.16} and \textcolor{black}{0.6}\Msun\ that didn't show any flares in their \tess\ lightcurves. These stars represent those that may be identified as preferable hosts of habitable planets, and selected for follow up observations to characterise exoplanetary atmospheres. 
The analysis of these stars is detailed in Sects.\,\ref{sec:method_tess} and \ref{sec:method_testing_models}.

\section{Methods} \label{sec:methods}
In this section we discuss the methods we used to detect flares in optical and FUV lightcurves, and calculate their energies. We also detail how we generated lightcurves from the FUV TIME-TAG data. 

\begin{table*}
    \centering
    \begin{tabular}{l|c|c|c|c|c|c|c|c}
     Name & $\mathrm{WL_{130}}$ & 
     \fuvonethirty\ & $\mathrm{PC_{130}}$ & Si IV & Si III & C II & C III & N V \\
     \hline
     9000\,K BB & $1.12\times10^{-3}$ & $2.52\times10^{-3}$ & $9.56\times10^{-4}$ & $3.80\times10^{-5}$ & $5.57\times10^{-5}$ & $2.66\times10^{-5}$ & $1.24\times10^{-5}$ &  
     $1.44\times10^{-5}$ \\ 
     \hline
     $\mathrm{FUV_{Model}}$/$\mathrm{FUV_{9000 BB}}$ & \\
     \hline
     Adjusted BB & 1.11 & 1.11 & 1.11 & 1.11 & 1.11 & 1.11 & 1.11 & 1.11 \\ 
     AD Leo Great Flare, 1 & 1.78 & 1.35 & 1.57 & 1.57 & 9.00 & 2.05 & 8.90 & 1.97 \\
     AD Leo Great Flare, 2 & 2.02 &  1.53 & 1.78 & 1.78 & 10.2 & 2.32 & 10.1 & 2.24 \\
     AD Leo Great Flare, 3 & 2.92 &  2.21 & 2.57 & 2.57 & 14.8 & 3.36 & 14.6 & 3.24 \\ 
     MUSCLES Model & 20.9  & 14.8 & 6.8 & 154 & 952 & 148 & 535 & 138 \\ 
    \end{tabular}
    \caption{Top row: The fraction of the 9000\,K blackbody bolometric energy emitted in each wavelength region. Bottom rows: The ratio of the energy emitted by each model and the 9000\,K blackbody for each wavelength region.}
    \label{tab:adjustment}
\end{table*}

\subsection{\tess\ Flare Detection and Energy Calculation} \label{sec:method_tess}
We used the same method as the one outlined in \citet{Jackman23} to detect stellar flares in the \tess\ lightcurves. We provide a brief outline here.  
This method first detrends the lightcurve by iteratively applying a median filter with a window size based on the modulation period in the \tess\ lightcurve. This step is done to remove quiescent modulation due to rotation, which results in starspots moving in and out of view to the observer. The method then searches for flares in the detrended lightcurve. Signals with three consecutive points above a set threshold are flagged as flare candidates. We set our threshold as 6 median absolute deviations (MADs) above the median of a given lightcurve. We visually inspected each flare candidate lightcurve to remove false positive signals.

We used the method outlined in \citet{Shibayama13} to calculate the energy of each flare in the \tess\ lightcurves. This method uses the observed flare amplitude, as measured from the lightcurve, to normalise the ratio between the flare and quiescent stellar spectrum within the observed bandpass. The bolometric flare energy is calculated by integrating the renormalised flare spectrum over wavelength and over the full duration of the flare. We first assumed the flare spectrum can be modelled by a 9000\,K blackbody, a common assumption based on atmospheric modelling of flare continuum fluxes \citep[][]{Hawley92} and later optical spectroscopy \citep[e.g.][]{Kowalski13}. This model has been widely used to calculate bolometric energies and rates of white-light flares \citep[e.g.][]{Gunther20}. This is despite studies showing flares showing a wide range of temperatures \citep[e.g.][]{Loyd18hazmat,Howard20}, due in part to our limited knowledge of how temperatures evolve and relate to other flare properties such as impulse. 
We determined the start and end time of each flare by eye to ensure that we integrated over the full duration of the flare, instead of between when it passed above and below the threshold used in our flare detection.   

We ran flare injection and recovery tests with the \tess\ lightcurves for each star in our sample to measure the efficiency of our detection method in terms of flare energy \citep[][]{Jackman23}. 
We measured the average recovery fraction $R(E)$ for each set of stars in our sample. We then used $R(E)$ to measure the average flare rate following \citet{Jackman21}. 
We measured flare rates using a Markov-Chain Monte Carlo method, using 32 walkers for 10,000 steps. We took the final 2000 steps of each run to sample the posterior distribution in our fitting of the flare rate. 

For stars in our optically quiet sample we used our measured recovery fractions to place limits on the maximum flare energies we would expect to detect in the \tess\ lightcurves. We calculated the energy that corresponded to an average detection efficiency of 99.7 per cent and used this to estimate the flare rate. We assumed that the occurrence of flares above this energy could be modelled with a Poisson distribution. We specified a flare rate that would give six flares during our \tess\ observations. A flare rate corresponding to one flare occurring has a 38 per cent chance of a non-detection. However, requiring six flares causes the probability of detecting zero flares to drop to 0.25 per cent. When combined with the energy limit calculated from our injection and recovery tests, this provides a conservative upper limit to the flare rate of these stars that enables us to better constrain whether the FUV activity of these stars can be predicted. Studies have previously measured values for the power law exponent $\alpha$ for the flare rates of low-mass stars between 1.4 and 2.5 \citep[see a review by][]{Ilin21}. 
However, this spread may depend on detection method. We chose a value of 2 for $\alpha$ to ensure our estimated maximum flare rate sat within this range.

\subsection{\hst\ Flare Detection and Energy Calculation}

Previous studies using \hst\ data have used different flare detection methods, ranging from visual inspection of individual lightcurves \citep[e.g.][]{Feinstein22} to the use of Gaussian process methods to search for consecutive outliers while modelling noise \citep[][]{Loyd18muscles}. We utilised a version of the method used for flare detection in \galex\ UV lightcurves created by \citet{Brasseur19}. This method was designed to search for short duration flare events in interrupted UV data and is efficient to run on large datasets. We ran our flare detection method on each of the FUV lightcurves generated in Sect.\,\ref{sec:data_hst_lc_generation}. We did this for more accurate comparison of our measured flare rates with those we will retrieve from flare injection and recovery tests in Sect.\,\ref{sec:method_testing_models}.

Lightcurves were first analysed for data points lying more than 3.5$\sigma$ above the lightcurve's global median. Here $\sigma$ refers to the Poisson uncertainty associated with each data point. Candidates were also required to have an adjacent point at least 2$\sigma$ above the lightcurve's global median, and the maximum value in the candidate lightcurve had to lie further from the global median than the minimum value in the lightcurve. The flare boundaries were where the candidate flux reached the lightcurve median, or the edge of the respective \hst\ visit, following \citet{Brasseur19}. We used this method to search for flares in each UV lightcurve for each source. For a single target with full FUV coverage, this resulted in eight flare searches. We searched each UV lightcurve individually to account for the different noise properties present for each integration region, something that impacts detection efficiency in our flare injection and recovery tests. Flare candidates were visually inspected to reject falsely flagged events.  

We used the flare boundaries measured in each lightcurves to define the start and end time of detected events. We used these timings in order to calculate flare energies. Energies were calculated following the method outlined in \citet{Loyd18muscles}. We took the median of the lightcurve in the 10 minutes preceding each candidate as a measure of the quiescent flux. This was subtracted from the candidate to isolate the flare-only emission. \textcolor{black}{During our analysis we identified two flares where a preceding flare had occurred within 10 minutes. In each case the overlap of the flare did not increase the median value beyond 1 $\sigma$ of the quiescent lightcurve outside of the overlapping region in time. We therefore elected to keep the automatically calculated median values for consistency with the rest of our sample. }
The flare-only emission was integrated over in time and multiplied by $4\pi d^{2}$, where $d$ is the distance to the star. This was performed for all candidates in each lightcurve. 

\begin{table*}
    \centering
    \begin{tabular}{l|c|c|c|c|c|c|c|c|c}
     Name & $\mathrm{WL_{130}}$ & 
     \fuvonethirty\ & $\mathrm{PC_{130}}$ & Si IV & Si III & C II & C III & N V  \\
     \hline
     9000\,K BB & $8.7^{+2.0}_{-1.8}$ & $6.9^{+1.7}_{-1.5}$ & $4.6^{+2.7}_{-2.2}$ & \textemdash\ & \textemdash\ & \textemdash\ & \textemdash\ & \textemdash\ & \\ 
     Adjusted BB & $7.84^{+1.80}_{-1.62}$ & $6.2^{+1.5}_{-1.4}$ & $4.14^{+2.43}_{-1.98}$ & \textemdash\ & \textemdash\ & \textemdash\ & \textemdash\ & \textemdash\ & \\ 
     AD Leo Great Flare, 1 & $4.89^{+1.12}_{-1.01}$ & $5.1^{+1.3}_{-1.1}$ & $2.93^{+1.72}_{-1.40}$ & $19^{+5.9}_{-4.8}$& $11^{+4.2}_{-3.5}$ & N/A & $50.56^{+25.84}_{-26.97}$ & N/A \\
     AD Leo Great Flare, 2 & $4.31^{+0.99}_{-0.89}$ & $4.5^{+1.1}_{-1.0}$ & $2.58^{+1.52}_{-1.24}$ & $17^{+5.2}_{-4.3}$ & $9.7^{+3.7}_{-3.1}$ & N/A & $44.55^{+22.77}_{-23.76}$ & N/A \\
     AD Leo Great Flare, 3 & $2.98^{+0.68}_{-0.62}$ & $3.1^{+0.8}_{-0.7}$ & $1.79^{+1.05}_{-0.86}$ & $12^{+3.6}_{-3.0}$ & $6.7^{+2.5}_{-2.2}$ & N/A & $30.82^{+15.75}_{-16.44}$ & N/A \\
     MUSCLES Model & $0.42^{+0.10}_{-0.09}$ & $0.5\pm0.1$ & $0.68^{+0.40}_{-0.32}$ & $0.19^{+0.06}_{-0.05}$ & $0.10^{+0.04}_{-0.03}$ & N/A & $0.84^{+0.43}_{-0.45}$ & N/A \\
    \end{tabular}
    \caption{UV energy correction factors (Sect.\,\ref{sec:method_testing_models}) for the partially convective field age M star sample. Each value is the factor one needs to apply to the FUV energy predicted by a corresponding model. N/A indicates a band where, due to too few detections, a flare occurrence rate could not be measured. The uncertainties correspond to the 16th and 84th percentiles in our fitting of the correction factors. \textcolor{black}{Dashed entries indicate where we have chosen not the provide correction factors from blackbody-based models for emission line features.}}
    \label{tab:correction_factors_pc}
\end{table*}

\begin{table*}
    \centering
    \begin{tabular}{l|c|c|c|c|c|c|c|c|c|c|c}
     Name & $\mathrm{WL_{130}}$ & 
     \fuvonethirty\ & $\mathrm{PC_{130}}$ & Si IV & Si III & C II & C III & N V \\
     \hline
     9000\,K BB & $99_{-31}^{+41}$ & $42\pm8$ & $32^{+20}_{-13}$ & \textemdash\ & \textemdash\ & \textemdash\ & \textemdash\ & \textemdash\ & \\ 
     Adjusted BB & $89.2^{+36.9}_{-27.9}$ & $38\pm7$ & $28.83^{+18.02}_{-11.71}$ & \textemdash\ & \textemdash\ & \textemdash\ & \textemdash\ & \textemdash\ & \\ 
     AD Leo Great Flare, 1 & $55.6^{+23.0}_{-17.4}$ & $31\pm6$ & $20.38^{+12.74}_{-8.28}$ &$303.18^{+213.38}_{-138.22}$ & N/A & $339.02^{+211.71}_{-145.37}$ & $385.39^{+171.91}_{-119.10}$ & $566.19^{+257.17}_{-240.88}$  \\
     AD Leo Great Flare, 2 & $49.0^{+20.3}_{-15.3}$ & $28\pm5$ & $17.98^{+11.24}_{-7.30}$ & $267.42^{+188.20}_{-121.91}$ & N/A & $299.57^{+187.07}_{-128.45}$ & $339.60^{+151.49}_{-104.95}$ & $497.94^{+226.17}_{-211.85}$ \\
     AD Leo Great Flare, 3 & $33.9^{+14.0}_{-10.6}$ & $19\pm4$ & $12.45^{+7.78}_{-5.06}$ & $185.21^{+130.35}_{-84.44}$ & N/A & $206.85^{+129.17}_{-88.69}$ & $234.93^{+104.79}_{-72.60}$ & $344.26^{+156.37}_{-146.46}$ \\
     MUSCLES Model & $4.7^{+2.0}_{-1.5}$ & $2.9\pm0.5$ & $4.71^{+2.94}_{-1.91}$ & $3.09^{+2.18}_{-1.41}$ & N/A & $4.70^{+2.93}_{-2.01}$ & $6.41^{+2.86}_{-1.98}$ & $8.08^{+3.67}_{-3.44}$ \\
    \end{tabular}
    \caption{UV energy correction factors for the fully convective field age sample. Note the increase relative to the partially convective sample. We discuss this further in Sect.\,\ref{sec:discussion_pc_fc}.}
    \label{tab:correction_factors_fc}
\end{table*}

\subsection{Testing Flare Models} \label{sec:method_testing_models}

We tested \textcolor{black}{six} flare models in this work. 
Each model describes both the optical and UV flare emission, and have been used in the literature to either calculate the bolometric energy or the UV emission from flares by calibration with optical flare lightcurves.  
These are the 9000\,K blackbody, the adjusted 9000\,K blackbody, the 1985 AD Leo Great Flare + 9000\,K blackbody and the MUSCLES flare model. Full descriptions of how each model is constructed and their use in the literature can be found in \citet{Jackman23}, who tested each one in the \galex\ NUV and FUV bands.

We tested the prediction of each model in the wavelength regimes listed in Tab.\,\ref{tab:line_emission} using the flare injection and recovery tests outlined in \citet{Jackman23}. We calculated the fraction of their energy each model emits in each FUV integration region. As we measured the flare rate of each sample in Sect.\,\ref{sec:method_tess} with bolometric energies calculated from the 9000\,K blackbody model, we also calculated the ratio of each model's FUV emission to that of the 9000\,K blackbody. This enabled us to adjust our measured flare rates from Sect.\,\ref{sec:method_tess} to the FUV prediction of each model accordingly. We used each fraction with the \tess\ bolometric flare rate to calculate a corresponding FUV rate for each sample.

We used the \citet{Davenport14} flare model to generate a grid of flares. These flares had amplitudes drawn from a uniform distribution between \textcolor{black}{0.1} and \textcolor{black}{10} times the quiescent lightcurve. They had $t_{1/2}$ durations drawn from a uniform distribution between \textcolor{black}{30 seconds and 5 minutes} \citep[e.g.][]{Loyd18muscles,Brasseur19,Jackman23}. We drew flares from this grid with energies according to the predicted FUV flare rate. We injected these flares into a quiescent, \textcolor{black}{flare-masked}, lightcurve and then ran the \hst\ flare detection algorithm to see which flares we could recover. We tested each model 10,000 times for each of the testing categories given above. We combined the COS and STIS generated lightcurves in this work.

\begin{figure}
    \centering
    \includegraphics[width=\columnwidth]{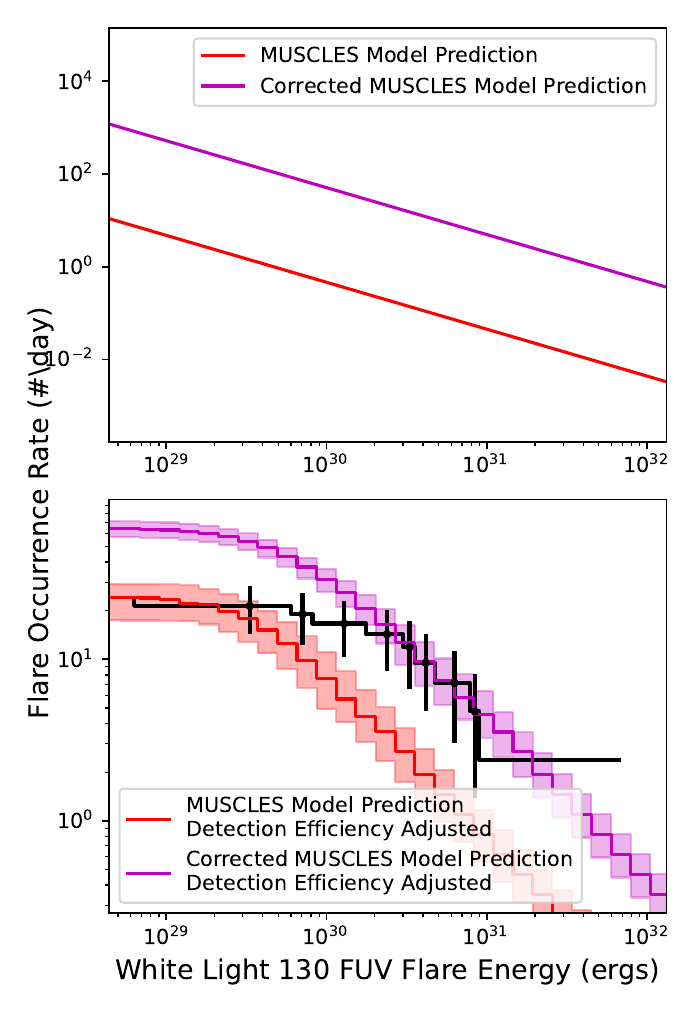}
    \caption{\textcolor{black}{Top: Predicted white-light 130 flare rates for the 40 Myr Tuc-Hor sample, using the MUSCLES model. The red is the flare rate obtained using by calibrating the model to the \tess\ flare rate, and the purple is the rate after applying the energy correction factor. 
    Bottom: The predicted FUV flare rates after performing flare injection and recovery tests with the \hst\ lightcurve, accounting for the detection efficiency in the \hst\ lightcurves. Also shown is the observed \hst\ FUV for the 40 Myr Tuc-Hor sample.}}
    \label{fig:flare_rate_example_1}
\end{figure}

We initially performed our flare injection and recovery tests using the originally predicted FUV flare rate for each model and integration region. However, some of these tests did not inject enough recoverable flares for us to study the discrepancy between the predicted and observed FUV flare rate. Those that did provided 
a qualitative measure of how well a model performed, but did not tell us by how much the FUV prediction of a model needs adjusting by to match the true flare emission. To determine this, we calculated FUV energy correction factors (ECFs) for each model. The ECF is the value a FUV model prediction (from extrapolating the \tess\ flare energy) needs to be multiplied by in order for the prediction to match FUV observations. 
This is because our observed average flare rates include selection effects related to the efficiency of our detection method (in particular for low energy flares) and the limited lengths of individual \hst\ visits. \textcolor{black}{An example of the energy correction factor in use is shown in Fig.\,\ref{fig:flare_rate_example_1}.} 
To measure energy correction factors for different models 
we performed a further series of injection and recovery tests \citep[][]{Jackman23}. We injected flare rates corresponding to increasing FUV energy fractions into each \hst\ lightcurve. We used FUV energy fractions corresponding to integer multiples of the energy fraction of the 9000\,K blackbody model. The results of our tests were used to construct a grid of recovered flare rates. We used an MCMC process to fit this grid to our measured average flare rate. Each step of the MCMC process sampled a randomly selected one of the 10,000 injection and recovery tests. We did this to incorporate uncertainties from our injection tests. 

To provide an initial estimate of the energy correction factor and speed up our analysis, we compared the UV flare rate predicted by each power law against the measured average flare rate for each sample. In this case the predicted UV flare rate did not take selection effects or different observing durations into account. However, comparing the initial predicted and measured flare rates provided a useful estimate of the required UV ECFs in each sample. We then specified a range of energy correction factors based around this estimate. 

We measured the average optical and UV flaring activity in Sect.\,\ref{sec:method_tess} and our model testing. We did this to try and mitigate changes in flare behaviour due to activity cycles. Previous studies \citep[e.g.][]{Jackman23} have used larger samples than those available here to average over the effects of activity cycles, or have benefited from simultaneous observations that remove this problem completely. Due to the small size of some of our samples (e.g. field age fully convective), there may be changes in the underlying stellar activity between our optical and FUV datasets. However, while the strength of optical and FUV emission lines have been observed to change during activity cycles \citep[e.g.][]{Loyd23}, changes in flare rates at the energies detected \tess\ have been tentative \citep[][]{Davenport20,Martin23}.  Therefore, we note the possible presence of these effects in our results, but add that without simultaneous optical and FUV observations we are limited in our testing of flare models. 

The results of our fitting are given in Tabs.\,\ref{tab:correction_factors_pc}, \ref{tab:correction_factors_fc} and \ref{tab:correction_factors_tuchor}. These factors are used in the same way as those provided in \citet{Jackman23} for the \galex\ NUV and FUV bandpasses. The FUV prediction of a chosen flare model should be multiplied by the relevant ECF in order to bring any subsequent predicted flare rate in line with observation.

\section{Results} \label{sec:results}
We detected \textcolor{black}{46} flares in our \wlonethirty\ lightcurves from our full sample of low-mass stars. The measured ECFs for each stellar sample are given in Tabs.\,\ref{tab:correction_factors_pc}, \ref{tab:correction_factors_fc}, \ref{tab:correction_factors_tuchor}, \ref{tab:correction_factors_tuchor_field} and \ref{tab:correction_factors_optically_quiet}. We discuss the results in the context of each flare model below.

\begin{table*}
    \centering
    \begin{tabular}{c|c|c|c|c|c|c|c|c}
     Name & $\mathrm{WL_{130}}$ & 
     \fuvonethirty\ & $\mathrm{PC_{130}}$ & Si IV & Si III & C II & C III & N V  \\
     \hline
     9000\,K BB & $104^{+47}_{-36}$ & $78^{+30}_{-24}$ & $32^{+12}_{-11}$ & \textemdash & \textemdash & \textemdash & \textemdash & \textemdash \\ 
     Adjusted BB & $93.69^{+42.34}_{-32.43}$ & $70^{+27}_{-22}$ & $28.83^{+10.81}_{-9.91}$ & \textemdash & \textemdash & \textemdash & \textemdash & \textemdash \\ 
     AD Leo Great Flare, 1 & $58.43^{+26.40}_{-20.22}$ & $58^{+22}_{-18}$ & $20.38^{+7.64}_{-7.01}$ & $1185^{+369}_{-357}$ & $28^{+9}_{-8}$ & $248^{+100}_{-80}$ & $130^{+47}_{-38}$ & $187.82^{+116.75}_{-81.22}$  \\
     AD Leo Great Flare, 2 & $51.49^{+23.27}_{-17.82}$ &$51^{+20}_{-16}$  & $17.98^{+6.74}_{-6.18}$ & $1045^{+326}_{-315}$ & $25^{+8}_{-7}$ & $219^{+88}_{-71}$ & $115^{+41}_{-33}$ & $165.18^{+102.68}_{-71.43}$ \\
     AD Leo Great Flare, 3 & $35.62^{+16.10}_{-12.33}$ & $35^{+14}_{-11}$ & $12.45^{+4.67}_{-4.28}$ & $724^{+226}_{-218}$ & $17^{+6}_{-5}$ & $151^{+61}_{-49}$ & $79^{+28}_{-23}$ & $114.20^{+70.99}_{-49.38}$ \\
     MUSCLES Model & $4.98^{+2.25}_{-1.72}$ & $5.3^{+2.0}_{-1.6}$ & $4.71^{+1.76}_{-1.62}$ & $12.1^{+3.8}_{-3.6}$ & $0.26^{+0.09 }_{-0.07}$ & $3.4^{+1.4}_{-1.1}$ & $2.2^{+0.8}_{-0.6}$ & $2.68^{+1.67}_{-1.16}$ \\
    \end{tabular}
    \caption{UV energy correction factors for the 40 Myr Tuc-Hor associated sample. Each value is the factor one needs to apply to the FUV energy predicted by a corresponding model. }
    \label{tab:correction_factors_tuchor}
\end{table*}

\begin{table*}
    \centering
    \begin{tabular}{l|c|c|c|c|c|c|c|c|c|c}
     Name & $\mathrm{WL_{130}}$ & 
     \fuvonethirty\ & $\mathrm{PC_{130}}$ & Si IV & Si III & C II & C III & N V \\
     \hline
     9000\,K BB & $11.0^{+2.3}_{-2.0}$ & $10.5^{+2.4}_{-2.3}$  &  $6.4^{+2.2}_{-2.1}$ & \textemdash\ & \textemdash\ & \textemdash\ & \textemdash\ & \textemdash\ \\ 
     Adjusted BB & $9.91^{+2.07}_{-1.80}$ & $9.46^{+2.16}_{-2.07}$ &  $5.77^{+1.98}_{-1.89}$ & \textemdash\ & \textemdash\ & \textemdash\ & \textemdash\ & \textemdash\ \\ 
     AD Leo Great Flare, 1 & $6.18^{+1.29}_{-1.12}$ & $7.78^{+1.78}_{-1.70}$ & $4.08^{+1.40}_{-1.34}$ & $23.50^{+4.90}_{-3.89}$ & $22.89^{+6.33}_{-6.00}$ & $37.73^{+14.32}_{-12.85}$ & $14.72^{+4.16}_{-3.48}$ & N/A \\
     AD Leo Great Flare, 2 & $5.45^{+1.14}_{-0.99}$ & $6.86^{+1.57}_{-1.50}$ & $3.60^{+1.24}_{-1.18}$ & $20.73^{+4.33}_{-3.43}$ & $20.20^{+5.59}_{-5.29}$ & $33.34^{+12.65}_{-11.36}$ & $12.97^{+3.66}_{-3.07}$ & N/A \\
     AD Leo Great Flare, 3 & $3.77^{+0.79}_{-0.68}$ & $4.75^{+1.09}_{-1.04}$ & $2.49^{+0.86}_{-0.82}$ & $14.36^{+3.00}_{-2.37}$ & $13.92^{+3.85}_{-3.65}$ & $23.02^{+8.74}_{-7.84}$ & $8.97^{+2.53}_{-2.12}$ & N/A \\
     MUSCLES Model & $0.53^{+0.11}_{-0.10}$ & $0.71^{+0.16}_{-0.16}$ & $0.94^{+0.32}_{-0.31}$ & $0.24^{+0.05}_{-0.04}$ & $0.22^{+0.06}_{-0.06}$ & $0.52^{+0.20}_{-0.18}$ & $0.24^{+0.07}_{-0.06}$ & N/A \\
    \end{tabular}
    \caption{UV energy correction factors for the field age counterpart to the 40 Myr Tuc-Hor sample. Each value is the factor one needs to apply to the FUV energy predicted by a corresponding model. }
    \label{tab:correction_factors_tuchor_field}
\end{table*}

\subsection{9000\,K Blackbody} \label{sec:results_9000}
The first model we tested was the canonical 9000\,K blackbody spectrum \citep[e.g.][]{Shibayama13}. As discussed in \citet{Jackman23} and Sect.\,\ref{sec:intro} this model is widely used in flare studies for calculating the bolometric energy of the footpoint emission. It has also been used to estimate the FUV emission of flares and their impact on exoplanetary atmospheres \citep[e.g.][]{Feinstein20}. We found that this model underestimated the line, pseudo-continuum and total FUV flare emission for all masses, ages and activity levels. The result for the line emission is not surprising, as this model is designed to only describe the continuum emission from the photospheric flare footpoints. \textcolor{black}{Consequently, we have chosen not to provide correction factors for emission line features where a model only provides blackbody continuum emission. This is to avoid errors when predicting FUV energies of individual emission lines that arise in atmospheric layers other than the lower chromosphere/upper photosphere.}

\subsection{Adjusted 9000\,K Blackbody}
The second model we tested was the adjusted 9000\,K blackbody. This model uses the 9000\,K blackbody model to calculate the bolometric energy and then multiplies this by a factor of \textcolor{black}{1.1} to include a contribution from line emission in the UV \citet{Osten15}. This model underestimated the FUV emission from line, pseudo-continuum and combined integration regions. This model performed better than the original 9000\,K blackbody, due to the extra flux contribution from line emission. However, the different UV ECFs between individual emission lines is indicative of the varying strength of responses for chromospheric and transition region lines. This is something we discuss further in Sect.\,\ref{sec:discussion_line}.

\subsection{1985 AD Leo Flare + 9000\,K Blackbody}
The third, fourth and fifth models combined the 9000\,K blackbody spectrum with the spectrum of the 1985 AD Leo Great Flare. The two parts of the model were joined by equating the energy in the U band. \textcolor{black}{\citet{Hawley91} noted that the short-wavelength (1150-2000\AA) spectrum was saturated during the flare at the peaks of emission lines and in the continuum beyond 1790\AA in the short-wavelength spectra. Estimates of line fluxes were made in their analysis using Gaussian fits to unsaturated regions, however they note an uncertainty of 50 per cent in their estimates due to the saturation.}  
This model has also been used to calculate the NUV emission for the abiogenesis zone. As in \citet{Jackman23}, we tested three versions of this model that used different values for the fraction of energy emitted by the 9000\,K blackbody in the U band. These values were \textcolor{black}{6.7, 7.6 and 11} per cent \citep[][]{Ducrot20,Gunther20,Glazier20}. \citet{Jackman23} found that each of these models emitted less NUV and FUV flux than the 9000\,K blackbody by itself. 
In contrast, our values in Tab.\,\ref{tab:adjustment} show that this model emits more FUV flux than the 9000\,K blackbody spectrum in the wavelength ranges we are studying. These regions are at shorter wavelengths than those covered by the \galex\ FUV bandpass, where the flux from the 9000\,K is reduced. In addition, the presence of emission lines increases the available flux in both the line-focused and combined emission lightcurves.

Despite this model having a greater amount of flux available in the FUV, each of the tested models still underestimated the observed FUV flare emission. However, this model did fare notably better for individual emission lines than the 9000\,K blackbody and adjusted blackbody models. This is due to the presence of realistic emission lines in the spectra. Yet, it still underestimated each region including the pseudo-continuum. This is likely related to joining the archival spectrum with the 9000\,K blackbody curve in the U band, and the assumed U band energy, issues noted by \citet{Jackman23}. This will have suppressed the flux from the UV spectrum. However, the different UV ECFs between the line, pseudo-continuum and combined lightcurves indicate that a simple scaling to correct for the assumed total U band energy will not bring the FUV predictions of this model in line with observations.

\begin{table*}
    \centering
    \begin{tabular}{l|c|c|c|c|c|c|c|c|c|c}
     Name & $\mathrm{WL_{130}}$ & 
     \fuvonethirty\ & $\mathrm{PC_{130}}$ & Si IV & Si III & C II & C III & N V \\
     \hline
     9000\,K BB & $4.30\pm1.0$ & $3.6\pm0.8$ & N/A & \textemdash\ & \textemdash\ & \textemdash\ & \textemdash\ & \textemdash\ \\ 
     Adjusted BB & $3.87\pm0.90$ & $3.24^{+0.72}_{-0.72}$ & N/A 
     & \textemdash\ & \textemdash\ & \textemdash\ & \textemdash\ & \textemdash\ \\
     AD Leo Great Flare, 1 & $2.42\pm0.56$ & $2.67^{+0.59}_{-0.59}$ & N/A 
     & $15.61^{+3.06}_{-2.74}$ & $12.78^{+4.56}_{-4.11}$ & $8.93^{+3.56}_{-2.68}$ & $10.9^{+4.7}_{-3.8}$ & N/A \\
     AD Leo Great Flare, 2 & $2.13\pm0.50$ & $2.35^{+0.52}_{-0.52}$ & N/A  
     & $13.76^{+2.70}_{-2.42}$ & $11.27^{+4.02}_{-3.63}$ & $7.89^{+3.15}_{-2.37}$ & $9.6^{+4.1}_{-3.4}$ & N/A \\
     AD Leo Great Flare, 3 & $1.47\pm0.34$ & $1.63^{+0.36}_{-0.36}$ & N/A 
     & $9.53^{+1.87}_{-1.67}$ & $7.77^{+2.77}_{-2.50}$ & $5.45^{+2.17}_{-1.64}$ & $6.6^{+2.8}_{-2.3}$ & N/A \\
     MUSCLES Model & $0.21\pm0.05$ & $0.24^{+0.05}_{-0.05}$ & N/A 
     & $0.16^{+0.03}_{-0.03}$ & $0.12^{+0.04}_{-0.04}$ & $0.12^{+0.05}_{-0.04}$ & $0.18^{+0.08}_{-0.06}$ & N/A \\
    \end{tabular}
    \caption{UV energy correction factors for the field age optically quiet sample. These should be considered lower limits. A value less than 1 indicates the model overestimated the FUV emission, and this prediction can be used to constrain the maximum FUV flare rate of optically quiet stars.} 
    \label{tab:correction_factors_optically_quiet}
\end{table*}
\subsection{MUSCLES Model}
The final model we tested was the one from \citet{Loyd18muscles}. This model uses a 9000\,K blackbody to model the total optical emission and the NUV continuum emission, and uses an empirically defined model spectrum for the FUV emission. The FUV emission model was developed using the energy budget of individual FUV emission lines observed during flares with \hst. 

As for \citet{Jackman23}, this model provided the best estimate of the FUV flare emission. We attribute this to the empirically defined FUV continuum that increases the flux above that from the 9000\,K blackbody model, and the increased line emission relative to other models. This is the only model that overestimated the amount of FUV flux in our tests. For the \textcolor{black}{partially convective field age, field age Tuc-Hor counterpart and field age optically quiet} samples this model could be used with the \tess\ lightcurves to provide an upper limit on the FUV activity of a system. 

\section{Discussion} \label{sec:discussion}
We tested the accuracy of the FUV emission of literature flare models calibrated using \tess\ observations. We did this in both the FUV continuum and emission lines for young and field age M stars. Our field age M stars had mass ranges of 0.37--0.6\Msun\ (M0-M2) and 0.1--0.29\Msun (M4-M5). We found that all but the MUSCLES model underestimated the FUV emission at all ages.

\subsection{Flaring from Partially vs Fully Convective Stars} \label{sec:discussion_pc_fc}
In Sect.\,\ref{sec:results} we presented the calculated UV ECFs for the field age partially and fully convective samples. We found that the fully convective stars required greater UV correction factors than the partially convective sample. We observed this in all our tested lightcurves and measured that \wlonethirty\ lightcurves (using the 9000\,K blackbody model) required correction factors of \textcolor{black}{$8.7^{+2.0}_{-1.8}$} and \textcolor{black}{$99^{+41}_{-31}$} for partially and fully convective stars respectively. 

An increase in UV ECFs between partially and fully convective M stars was previously observed by \citet{Jackman23}, using \galex\ NUV and FUV photometry. 
They suggested that the increase in the \galex\ NUV correction factors could be explained by an increase in the average flare temperature for fully convective stars. \citet{Jackman23} calculated  pseudo-continuum blackbody temperatures of 11,500 and 15,800\,K were required to explain the optical and \galex\ NUV flare emission from partially and fully convective stars respectively. However, these temperatures decreased to 9000\,K and 10,700\,K when the Balmer continuum was taken into account. 
However, the blackbody spectra associated with the required flare temperatures underestimated the observed \galex\ FUV flare emission for the fully convective stars. 
One explanation for this was that the models tested did not sufficiently account for the FUV line and enhanced continuum emission. Our results show that even models that include the flux from individual emission lines underestimate the energy released by flares, something we discuss further in Sect.\,\ref{sec:discussion_line}.

\begin{figure}
    \centering
    \includegraphics[width=\columnwidth]{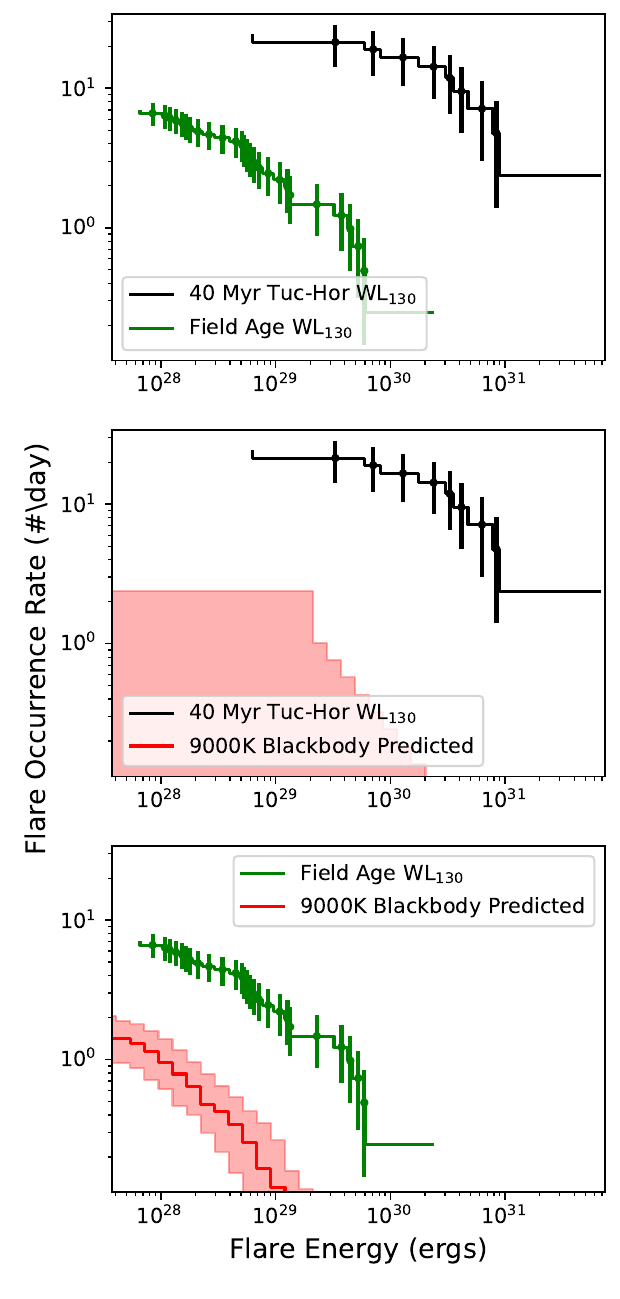}
    \caption{Comparison of the \wlonethirty\ flare rate for our 40 Myr Tuc-Hor sample and the field age sample. These samples have the same range of masses. Top: Comparison of the measured average flare rates. Note the decrease in measured occurrence rate and energy, something previously identified for this 40 Myr sample and field age stars by \citet{Loyd18hazmat}. Middle: Comparison of the detected FUV flare rate with the prediction of the 9000\,K blackbody model. This model doesn't predict enough retrievable flares for us to measure a flare rate, resulting in the shaded region being an upper limit. Bottom: The same as the middle plot, but for the field age sample. Note the reduced difference between the predicted and observed FUV flare rates for the field age sample, a sign of flare model accuracy increasing with age.}
    \label{fig:muscles_compare}
\end{figure}

These results lend weight to previous suggestions that optical flare emission observed by wide-field surveys should not be considered as a direct tracer of the FUV emission of flares. \textcolor{black}{\citet{Feinstein22} noted that the flare rate of AU Mic was likely higher in the FUV than in the optical, which they attributed to lower energy flares being easier detect in the FUV than at longer wavelengths. }Simultaneous optical and FUV flare observations by \citet{MacGregor21} showed the optical emission trailed the FUV, and that the relative amplitudes could not be explained by a single pseudo-continuum blackbody emitter. On the Sun, optical emission from flares has been detected from both low energy C class and high energy X class events \citep[][]{Hudson06} and has been suggested as a common feature of all flares \citep[][]{Jess08}. However, other studies have found that the occurrence rate of a white-light counterpart appears to increase with increasing energy, with \citet{Song18} and 
\citet{Castellanos20} measuring occurrence rates of 8.5-10.5 per cent for C-class flares and 94.4-100 per cent for X-class flares. \citet{Watanabe17} found evidence that the presence of a white-light counterpart (to a hard X-ray flare) depended on the rate of energy deposition, with short duration flares occurring in regions with stronger magnetic fields being more likely to generate white-light enhancements. If these results hold for low-mass stars, then we may only detect optical emission from flares with energy deposition rates above some limit that results in the heating of lower atmospheric layers, either directly or through a strong enough chromospheric condensation. If the FUV emission is associated with heating in the upper chromosphere from the initial energy deposition \citep[e.g. creating a chromospheric condensation;][]{Froning19,MacGregor21}, then we would expect to see FUV flares more regularly. \textcolor{black}{A change in the chromospheric density or structure of 
fully convective M stars relative to their partially convective counterparts \citep[e.g.][]{Schmidt15}  
could help explain a higher rate of FUV-only flares and thus UV ECFs.} We measured overlapping FUV energy ranges for the partially and fully convective samples. However our fitted \tess\ power laws give lower optical flare energies for the fully convective sample at the corresponding rates. If the energies required to trigger a detectable white-light response are higher for fully convective stars, then this could explain the increased UV correction factors for the fully convective sample.

\subsection{A Need for Age Dependent Flare Models?} \label{sec:discussion_age}

In Sect.\,\ref{sec:results} we tested our flare models for 40 Myr old M0-M2.3 stars in the Tuc-Hor association, and for field age stars in the same mass range. We found that the UV ECFs decreased between the 40 Myr and field age sample. Our measured rates for the \wlonethirty\ lightcurves are shown in Fig.\,\ref{fig:muscles_compare}, along with the predictions of the 9000\,K blackbody model for each sample. 

Younger stars having higher UV ECFs may point towards a change in the optical/FUV energy partition with age. This would be particularly important for studies of the evolution of exoplanet atmospheres, which simulate the XUV irradiation received by rocky planets over time \citep[e.g.][]{King21,Ketzer23}. One possibility is that younger stars exhibit higher average flare temperatures than their field age counterparts, resulting in greater than expected FUV emission. Our sample includes the ``Hazflare'' studied by \citet{Loyd18hazmat}, who measured a temperature of 15,500\,K across the impulsive phase that included the rise, peak and initial decay. Two-colour photometry of low-mass stars in open clusters can help constrain the evolution of flare temperatures with age. Another reason for a decrease in the UV ECFs may be a change in the number of FUV-only flares studied in each sample, similar to the difference between partially and fully convective field stars discussed in Sect.\,\ref{sec:discussion_pc_fc}. 

Another possibility is that a change in the UV ECF is due to the decreasing FUV energy of the flares we are probing. We measured \fuvonethirty\ and \wlonethirty\ energies approximately 410 and 300 times greater respectively for the 40 Myr sample than for the field age sample at the same flare rate. We note that \textcolor{black}{during the HAZMAT programme study \citep[][]{Shkolnik14} of Tuc-Hor associated stars, } \citet{Loyd18hazmat} \textcolor{black}{found that for a flare rate of 5 per day, the 40 Myr Tuc-Hor sample exhibited flares 100-1000 times more energetic than the }
field age M stars from the MUSCLES survey. 
Previous works have noted a difference in the slopes of the optical and UV flare rates \textcolor{black}{for individual stars and stellar samples}, resulting in a divergence at higher energies. \citet{Paudel21} measured the optical and NUV flare rates of the M4 star EV Lac with simultaneous \tess\ and \swift\ photometry. They measured a shallower rate in the NUV than the optical. \citet{Brasseur23} observed similar behaviour when comparing contemporaneous \kepler\ optical and \galex\ NUV flare observations. They found that flares with higher energies exhibited a higher NUV to optical flux ratio, pointing towards the flare SED changing with energy. \citet{Jackman23} also identified similar behaviour when comparing \galex\ NUV flare rates of low-mass stars with those predicted by \tess\ and flare models. As we were studying higher energy optical and FUV flares for the 40 Myr sample than the field age sample, a change in the optical/UV flux ratio with flare energy could explain our results \textcolor{black}{for the two age samples. While previous studies have found that the ratio between the FUV flare activity and quiescent emission stays constant with age \citep[e.g.][]{Loyd18hazmat}, changes with age will result in us probing different flare energies for the same stellar mass ranges. Consequently, if the FUV and optical flare rates exhibit different slopes \citep[e.g.][]{Brasseur23,Jackman23}, any divergence between the optical and FUV flare rate will be greater for younger ages. This will result in the measurement of higher energy correction factors for younger ages.}

Our results provide a useful test for activity evolution models that describe how flare rates change with time. An example of such a model is the one from \citet{Davenport19}, which was calibrated using \kepler\ photometry of flare stars with ages from gyrochronology. This model predicts that between 40 Myr and 5 Gyr the energy corresponding to a rate of 5 flares per day for an \textcolor{black}{0.5\,\Msun star } decreases by 38 per cent. Our results shows that this model underestimates the change in FUV emission, with an exception for Si III. Our results also show that the change with age is not uniform, with the correction factors for the pseudo-continuum and line emission changing by different amounts. We must acknowledge that the changes in our measured UV correction factors are likely driven by both changes in the optical and the UV.

\subsection{Optically Quiet, but FUV Loud}

\begin{figure}
    \centering
    \includegraphics[width=\columnwidth]{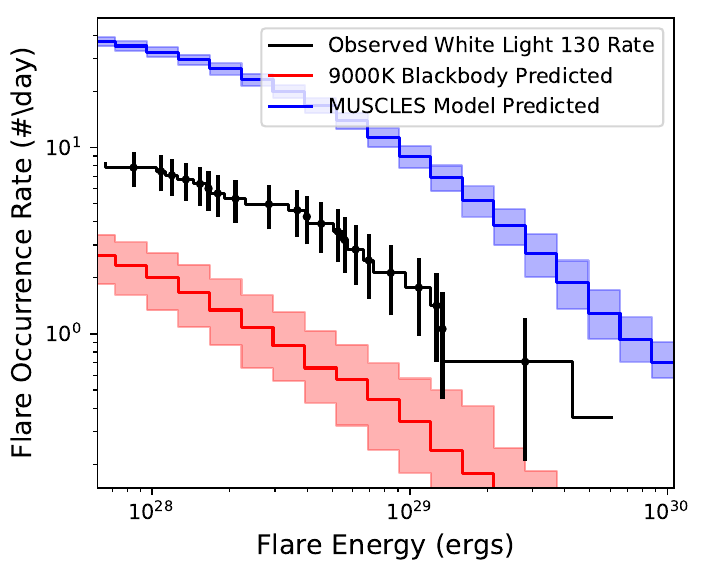}
    \caption{The \wlonethirty\ FUV flare rate for optically quiet low-mass stars. These stars exhibit no flares in their \tess\ 2-minute cadence lightcurves. The black line is the observed flare rate, showing our sample flares with an FUV energy above $10^{29}$ ergs on average twice a day. The red and blue regions are the predictions of the 9000\,K blackbody and MUSCLES models calibrated to the upper limit of possible optical flare rates. The MUSCLES model overestimates the FUV flare rate, constraining the maximum FUV activity of optically quiet low-mass stars.}
    \label{fig:optically_quiet}
\end{figure}

In Sect.\,\ref{sec:method_testing_models} we investigated the FUV activity of optically quiet stars. These stars exhibit no detectable flaring activity in their \tess\ lightcurves. This has led studies to highlight these and similar stars as amenable for habitability and as targets for atmospheric characterisation. However, \citet{Loyd20} detected FUV flares from the optically quiet star \textcolor{black}{GJ 887}, showing that these sources can still exhibit FUV flaring behaviour that may not be predictable by the optical observations alone. In Sect.\,\ref{sec:data_hst_lc_generation} we were able to confirm this result by measuring average FUV flare rates for a sample of optically quiet low-mass stars. Fig.\,\ref{fig:optically_quiet} shows the measured average FUV flare rate for the optically quiet sample, showing they flared with an energy above \textcolor{black}{$1\times10^{29}$ ergs} approximately once per day. 

The measurement of FUV flare rates for a sample of optically quiet stars calls into question the use of optical lightcurves in predicting high energy stellar activity. If the optical lightcurve can not be used to rule out FUV flare activity above some energy or rate, then studies of exoplanets will not be able to dismiss FUV emission from their atmospheric evolution modelling. To test this, we used our \tess\ flare injection and recovery tests in Sect.\,\ref{sec:method_testing_models} to characterise which flare energies we could detect in the optically quiet lightcurves, measuring the energy at which we would detect 99.7 per cent of flares. We then calculated a maximum flare rate based on this energy. This rate was considered an upper limit, above which we would confidently expect to detect one flare. We used this flare rate to estimate minimum FUV ECFs for the optically quiet sample, which are shown in Tab.\,\ref{tab:correction_factors_optically_quiet}. These values are considered minimum correction factors due to our use of the upper limit of the \tess\ flare rate. 

We measured that all models except for the MUSCLES model had FUV ECFs greater than 1. This indicates that even when we constrain the maximum optical flare rate, these models will still underestimate the FUV activity of low-mass stars. The MUSCLES model has ECFs less than 1, in particular for the 
\fuvonethirty\ and line emission. When used with the MUSCLES model the maximum flare rate derived from our \tess\ injection and recovery tests overestimates the FUV activity. This effect can be seen in Fig.\,\ref{fig:optically_quiet} and suggests that this model could be used to constain the maximum FUV flare rate of an optically quiet star, but will still not provide an accurate estimate to the true FUV activity.

\subsection{The Effect of Emission Lines on FUV Predictions} \label{sec:discussion_line}
The benefit of using spectroscopic data is that we were able to isolate the line and pseudo-continuum emission in our analysis. We measured decreases in the UV ECFs for all samples when we isolated the pseudo-continuum emission, relative to the lightcurves that included both continuum and line emission. In some cases the UV ECF changed by over 50 per cent between the \wlonethirty\ and pseudo-continuum lightcurves, highlighting the contribution from emission lines to the FUV flux of flares. 
\citet{Hawley03} measured emission lines contributing between \textcolor{black}{34} and \textcolor{black}{40} per cent of the FUV energy budget of flares detected from AD Leo, while \citet{Feinstein22} measured individual lines contributing up to 21 per cent. 
\citet{France16} discussed how the contribution of individual lines to the FUV energy budget likely depends on formation temperature, with lines forming near Si III (in terms of temperature) responding most strongly to the flare energy deposition. \citet{Feinstein22} identified similar behaviour in their analysis of FUV flares from the active young M star AU Mic. They found that C II and C III contributed the largest fractions of their measured FUV flux, followed by Si III. These lines form in the upper chromosphere and transition region, the layers at which energy deposition from reconnection events is expect to occur. \citet{France16} and \citet{Loyd18muscles} noted that the line response drops off for cooler (lower chromospheric layers) and higher (coronal) temperatures.

It is not surprising then that our worst performing model is the 9000\,K blackbody model, something noted in Sect.\,\ref{sec:results_9000}, given that is designed to describe the continuum emission at the photospheric footpoints. Other models that incorporate the contribution from line emission fare better. In contrast to the other models, the MUSCLES model overestimates the emission in individual lines for field age stars.
This model used observations from \citet{Hawley03} to fix the bolometric energy of the 9000\,K blackbody continuum relative to the Si IV flux.  The overestimation for field age stars may be due to a flux jump as the model switches between the 9000K blackbody and empirically calibrated FUV emission. This jump introduces a new level of continuum emission (akin to the Balmer jump), increasing the FUV emission above that of the other models. However, it still underestimates the flux in both the line and pseudo-continuum regions for some of our samples. This suggests that the energy ratios between continuum and line emission also change with age and/or flare energy. Interestingly the AD Leo Great Flare models, that utilise FUV flare spectra, still underestimate FUV flare line emission. \textcolor{black}{This may be due to the reconstruction of emission lines by \citet{Hawley91}, due to the saturation of line peaks in the IUE short wavelength spectrum.}
This may also be due to the use of the U band and assumed energy of $10^{34}$ ergs to combine the 9000\,K blackbody and optical+UV spectra, which was noted to suppress the UV emission by \citet{Jackman23}.
Another element is the use of \tess\ photometry to calibrate each model. The \tess\ photometry probes the photospheric emission. As mentioned above the FUV emission lines are formed in the upper chromosphere and transition region. These regions will have different responses and temporal morphologies to the optical emission,
limiting its use to calibrate models. This is apparent in our measured UV ECFs, in particular the decrease in ECF value between the total (line and continuum) and pseudo-continuum lightcurves. While previously noted for individual flares \citep[e.g.][]{MacGregor21}, our results indicate that these effects impact predictions of the average flare behaviour also.

\subsection{Comparison with \galex\ FUV}
In Sect.\,\ref{sec:method_sample} we specified the mass ranges of our field age samples to be the same as those used by \citet{Jackman23}. \citet{Jackman23} used \galex\ broadband photometry to test the \galex\ FUV accuracy of flare models. They found that the 9000\,K blackbody model underestimated the \galex\ FUV emission of flares from fully convective field age M stars by a factor of $30.6\pm10.0$. In Sect.\,\ref{sec:results} we found that the 9000\,K blackbody model underestimated the integrated \fuvonethirty\ flare emission of field age fully convective stars by a factor of $42\pm8$. While these values are consistent within $2\sigma$, the \hst\ \fuvonethirty\ ECF is greater than that calculated for the \galex\ bandpass. We attribute this to the different wavelength coverage in each study. The \galex\ FUV bandpass covers wavelengths of 1344--1786\AA, while our white-light lightcurves cover groups of wavelengths between 1173.65 and 1362.70\AA. The lightcurves in this study include flare emission at shorter wavelengths, where the 9000\,K blackbody continuum used by many models is reduced in strength. In addition, flares with temperatures greater than 9000\,K would emit greater amount continuum emission at bluer wavelengths covered in this work. The difference due to the covered wavelength is also reflected in our pseudo-continuum emission correction factors. The UV ECF of $32^{+20}_{-13}$ is similar in value to the \galex\ FUV factor of $30.6\pm10.0$, but this lightcurve is designed to remove the contribution from emission lines.
The combined results of these works highlight that for field age fully convective stars, the UV predictions of optically calibrated flare models become progressively worse as we move to shorter wavelengths.

\section{Conclusions} \label{sec:conclusions}
We used optical and archival FUV observations of low-mass stars to test the UV predictions of literature flare models. These flare models had previously been tested using \galex\ NUV and FUV photometry by \citet{Jackman23}. We used the archival \hst\ TIME-TAG spectroscopy to test model predictions for continuum and line emission separately. 

We found that literature flare models underestimated the FUV emission of fully convective stars more than for partially convective stars. This was in line with previous results in the NUV, suggesting that the energy fractionisation of flares changes with the transition to a fully convective interior. We discussed that this may be due to an increased presence of UV flares that have non-detectable, or only weakly detectable, optical counterparts. However, we were limited in our analysis by the non-simultaneous nature of our observations. Future studies with further simultaneous optical and FUV observations will allow us to better constrain the relative occurrence of flares from these two wavelength regimes for low-mass stars. 

We also used archival observations of 40 Myr stars associated with the Tuc-Hor association to test whether the accuracy of flare models changes with age. By comparing UV correction factors for the 40 Myr sample and field age stars in the same mass regime, we identified that models underestimate FUV emission less severely as stars age. However, we found that the MUSCLES model overestimated the FUV emission of flares for field age stars.

\section*{Acknowledgements}
\textcolor{black}{The authors thank the anonymous referee for their helpful comments and role in improving this manuscript. }
JAGJ thanks Chase Million and Scott W. Fleming for their helpful discussions relating to this work. 
This research has made use of the SVO Filter Profile Service (http://svo2.cab.inta-csic.es/theory/fps/) supported from the Spanish MINECO through grant AYA2017-84089. 
JAGJ \textcolor{black}{and ES} acknowledge support from grant HST-AR-16617.001-A from the Space Telescope Science Institute, which is operated by the Association of Universities for Research in Astronomy, Inc., under NASA contract NAS 5-26555. 

This paper includes data collected by the \tess\ mission, which is publicly available from the Mikulski Archive for Space Telescopes (MAST). Funding for the \tess\ mission is provided by NASA's Science Mission directorate. This research was supported by NASA under grant number \textcolor{black}{80NSSC22K0125} from the \tess\ Cycle 4 Guest Investigator Program.

\section*{Data Availability}
All data used in this work is publicly available on MAST.



\bibliographystyle{mnras}
\bibliography{main} 








\bsp	
\label{lastpage}
\end{document}